\documentclass[twocolumn,byrevtex,prb]{revtex4}

\usepackage{graphicx}

\begin{document}

\title{Optical excitations in a one-dimensional Mott insulator}

\author{Eric Jeckelmann}
\affiliation{Fachbereich Physik, Philipps-Universit\"{a}t, 
D-35032 Marburg, Germany}

\date{\today}

\begin{abstract}
The density-matrix renormalization-group (DMRG) method
is used to investigate optical excitations in the Mott
insulating phase of a one-dimensional extended Hubbard model.
The linear optical conductivity is calculated using the dynamical
DMRG method and the nature of the lowest optically excited states 
is investigated using a symmetrized DMRG approach.
The numerical calculations agree perfectly with field-theoretical 
predictions for a small Mott gap and analytical results 
for a large Mott gap
obtained with a strong-coupling analysis.
Is is shown that four types of optical excitations exist in this Mott
insulator:
pairs of unbound charge excitations, excitons, excitonic strings, and 
charge-density-wave (CDW) droplets.
Each type of excitations dominates the low-energy optical spectrum
in some region of the interaction parameter space
and corresponds to distinct spectral features:
a continuum starting at the Mott gap (unbound charge excitations),
a single peak or several isolated peaks below the Mott gap (excitons 
and
excitonic strings, respectively), and a continuum 
below the Mott gap (CDW droplets).
\end{abstract}


\maketitle

\section{Introduction \label{sec:intro}}

In recent years, 
various quasi-one-dimensional materials, such as conjugated 
polymers,~\cite{kiess,sari} 
organic charge-transfer salts,~\cite{farges,bourbonnais}
Cu oxides,~\cite{hasan} and Ni halides,~\cite{kishida} have been 
extensively studied 
because of their unusual
optical properties and their potential application in modern
optical technology.
In first approximation these materials can be described
as one-dimensional strongly correlated electron systems
with half-filled bands.
The electron-electron interaction drives such a system into a
Mott insulating ground state~\cite{mott,florianbook} and dominates
low-energy excitations.
Therefore, the
optical properties of one-dimensional Mott insulators
are currently a topic of great interest.

Despite many theoretical studies, our knowledge of these systems 
is still fragmentary because of the difficulties associated with the 
investigation of strongly correlated systems.
For many years, numerical exact diagonalization of small
systems was the only method providing reliable information on
excited states in correlated electron 
systems.~\cite{guo,diago}
Recently, however, the linear optical conductivity and exciton 
properties of one-dimensional Mott insulators have been calculated
analytically for an infinite system in the limit of a small
Mott gap~\cite{JGE,controzzi,EGJ} and of a large 
Mott gap.~\cite{EGJ,florian,gallagher,stephan,mizuno,barford}
Moreover, recent developments
of the density-matrix renormalization group (DMRG)
method~\cite{steve,dmrgbook} allows one to calculate excited states
and dynamical response functions numerically
in large systems and with
an accuracy comparable to exact diagonalizations.~\cite{JGE,jeckel}

A paradigm of a one-dimensional Mott insulator is the
extended Hubbard model with hopping integral $t$, on-site repulsion $U$,
and nearest-neighbor repulsion $V$ at half filling.
Although this model has been widely studied, its properties are still
poorly understood in the thermodynamic limit. 
It is known~\cite{hirsch}
that the system is a Mott insulator in a large region 
of the parameter space $(U,V)$ which is physically relevant for the
Coulomb repulsion between electrons $(U > V \geq 0)$.
The precise ground-state phase diagram has only recently be determined
using DMRG.~\cite{jeckel2}
In the special case of the Hubbard model~\cite{hubbard}
($V=0, U> 0$) 
the ground state is known exactly to be a Mott 
insulator~\cite{lieb} and the optical conductivity
has been calculated.~\cite{JGE}
In principle, the optical properties of the Mott insulating phase
are also known for $V>0$
at weak coupling~\cite{controzzi,EGJ} and at strong
coupling,~\cite{EGJ,florian,gallagher,stephan,mizuno}
but the range of validity of these results is not clear a priori.
Results for intermediate coupling and close to the Mott phase boundary
$U\approx 2V$ are scarce
and the absence of finite-size-effect analysis often hinders their 
interpretation.~\cite{stephan,shuai,kancharla,tomita}

In this paper I present an accurate and comprehensive investigation
of the linear optical conductivity $\sigma_1(\omega)$ and the 
excited states contributing to $\sigma_1(\omega)$ in the
Mott insulating phase of the one-dimensional extended Hubbard model
at half filling. 
An efficient symmetrized DMRG~\cite{ramasesha} and the recently 
developed dynamical DMRG (DDMRG)~\cite{JGE,jeckel}
are used to calculate the optically excited states and 
the linear optical conductivity $\sigma_1(\omega)$ 
on large lattices.
The comparison of numerical results
for $\sigma_1(\omega)$ with field-theoretical~\cite{controzzi,EGJ} 
and strong-coupling~\cite{florian}
predictions confirms both the great accuracy of DDMRG and the
wide validity range of both analytical methods.
I have found that four types of excitations
determine the optical properties in the Mott insulating phase:
pairs of unbound charge excitations, excitons, excitonic strings,
and charge-density-wave (CDW) droplets.
I will show that each type of excitations dominate the low-energy 
spectrum
in a particular region of the parameter space $(U,V)$
and exhibits a distinct optical spectrum $\sigma_1(\omega)$.

The model, the linear optical conductivity $\sigma_1(\omega)$,
and the relevant symmetries are introduced in detail in the
next section. 
In Sec.~\ref{sec:methods} the numerical methods are briefly 
presented, then the estimation of DMRG truncation errors and 
finite-size effects are discussed.
In Sec.~\ref{sec:results} I describe the four different types 
of excitations which contribute to the linear optical conductivity
and the corresponding optical spectra in the various
interaction regimes from the limit of a large Mott gap
to the limit of a small Mott gap.
The final section contains the conclusion.

\section{Model \label{sec:model}}

The one-dimensional extended Hubbard model is defined by the
Hamiltonian
\begin{eqnarray}
\hat{H} 
&=& -t \sum_{l;\sigma} \left( \hat{c}_{l,\sigma}^+\hat{c}_{l+1,\sigma} 
+ \hat{c}_{l+1,\sigma}^+\hat{c}_{l,\sigma} \right) \nonumber \\
&&+ U \sum_{l} \left(\hat{n}_{l,\uparrow}-\frac{1}{2}\right)
\left(\hat{n}_{l,\downarrow}-\frac{1}{2}\right)  \nonumber \\
&& + V \sum_{l}(\hat{n}_l-1)(\hat{n}_{l+1}-1)  \; .
\label{hamiltonian}
\end{eqnarray}
It describes electrons with spin 
$\sigma=\uparrow,\downarrow$ 
which can hop between neighboring sites. Here $\hat{c}^+_{l,\sigma}$,
$\hat{c}_{l,\sigma}$ are creation and annihilation operators for
electrons with spin $\sigma$ at site $l$, $\hat{n}_{l,\sigma}=
\hat{c}^+_{l,\sigma}\hat{c}_{l,\sigma}$ are the corresponding density
operators, and $\hat{n}_l=\hat{n}_{l,\uparrow}+\hat{n}_{l,\downarrow}$.
The hopping integral $t > 0$ gives rise to a 
a single-electron band of width $4t$.
The Coulomb repulsion is
mimicked by a local Hubbard interaction $U$, 
and a nearest-neighbor interaction $V$. 
The physically relevant parameter regime is $U > V \geq 0$.
The number of electrons equals the
number of lattice sites $N$ (half-filled band). 
This system is in a Mott insulating phase for 
$V < V_c(U) \approx U/2$~(Ref.\onlinecite{hirsch}).
Precise values of the Mott phase boundary $V_c(U)$ are given in 
Ref.~\onlinecite{jeckel2}.

Note that the chemical potential is chosen in such a way 
that the Hamiltonian~(\ref{hamiltonian})
explicitly exhibits a particle-hole symmetry.
This Hamiltonian has two other discrete symmetries which are useful
for optical excitation calculations: a spin-flip symmetry
and a spatial reflection symmetry (through
the lattice center). 
Therefore, each eigenstate has
a well-defined parity under charge conjugation
($P_{c} = \pm 1$) and spin flip ($P_{s} = \pm 1$), and
belongs to one of the two irreducible
representations, $A_{g}$ or $B_{u}$, of a   
one-dimensional lattice reflection 
symmetry group.  

Spectroscopy with electromagnetic radiation
is a common experimental probe of solid-state 
materials.~\cite{kuzmany}
The linear (one-photon) optical absorption is proportional 
to the real part $\sigma_1(\omega)$ of the optical conductivity.
For $\omega \neq 0$, $\sigma_1(\omega)$
is related to the imaginary part of the current-current 
correlation function by
\begin{equation}
\sigma_1(\omega) =\frac{\text{Im}\{\chi(\omega)\}}{\omega} \; .
\label{sigma}
\end{equation}
For $\omega \geq 0$ the current-current
correlation function is given by
\begin{eqnarray}
\chi(\omega>0) 
&=& - \frac{1}{Na} \left \langle 0 \left |\hat{J} 
\frac{1}{E_0-\hat{H}+\hbar\omega+i\eta}\hat{J} \right |0 \right \rangle 
\nonumber \\
&=& - \frac{1}{Na} \sum_n 
\frac{ |\langle 0 | \hat{J} | n\rangle|^2}{\hbar\omega -(E_n-E_0) 
+i\eta} \; ,
\label{currentcorrel}
\end{eqnarray}
where $a$ is the lattice spacing.  
Here, $|0\rangle$ is the ground state of the
Hamiltonian $\hat{H}$, $|n\rangle$ are excited states of $\hat{H}$,
and $E_0$, $E_n$ are their respective eigenenergies. 
Although $\eta=0^+$ is infinitesimal,
a finite value may be used to broaden the resonances at 
$\hbar\omega=E_n-E_0$ and to reduce finite-size effects.
$\hat{J}$ is the current operator 
\begin{equation}
\hat{J}=\frac{iaet}{\hbar}  
\sum_{l;\sigma} \left( \hat{c}_{l,\sigma}^+\hat{c}_{l+1,\sigma} 
- \hat{c}_{l+1,\sigma}^+\hat{c}_{l,\sigma} \right) \; ,
\label{current}
\end{equation} 
where $-e$ is the charge of an electron. 
I set $a=e=\hbar = 1$ throughout, and $t=1$ is used
in figures showing the optical conductivity,
which means that $\sigma_1(\omega)$ is given in units
of $e^2a/\hbar$ and $\omega$ in units of $t/\hbar$.

We note that the current operator is invariant under the 
spin-flip transformation but antisymmetric under 
charge conjugation and spatial reflection. 
Therefore, if the ground state $|0\rangle$ belongs to
the symmetry subspace $A_{g}^{+} \equiv (A_{g},P_{c},P_{s})$, 
only excited states
$|n\rangle$ belonging to the symmetry subspace
$B_{u}^{-} \equiv (B_{u},-P_{c},P_{s})$ contribute to the 
optical conductivity.
According to selection rules, the matrix element 
$\langle0|\hat{J}|n\rangle$ vanishes if $|n\rangle$
belongs to another symmetry subspace.
In this paper, the excitation energy $E_n-E_0$ of the lowest
eigenstate with a non-zero matrix element $\langle0|\hat{J}|n\rangle$
is called the optical gap $E_{\text{opt}}$.

\section{Numerical methods \label{sec:methods}}

DMRG~\cite{steve,dmrgbook}
is known to be a very accurate method for one-dimensional
quantum systems with short-range interactions such as
the extended Hubbard Hamiltonian~(\ref{hamiltonian}).
In this work I use three different DMRG techniques to calculate
ground states, excited states, and dynamic response functions.
All three techniques are based on the finite system DMRG algorithm.

First, the usual ground-state DMRG method is used to calculate
the ground state $|0\rangle$ 
for  a fixed number $N_{\sigma}$ of electrons of each 
spin $\sigma$.
This method provides the ground-state energy 
$E_0(N_{\uparrow},N_{\downarrow})$ and allows us 
to calculate ground-state expectation values
$\langle 0 | \hat{O} | 0 \rangle$
for various operators $\hat{O}$, such as static correlation functions.
The (Mott) gap $E_{M}$ in the single-particle density of 
states of a Mott insulator can also be obtained using this approach.
At half filling ($N_{\sigma} = N/2$) the Mott gap is simply given by
\begin{equation}
E_{M} = 2 \ [ \ E_0(N_{\uparrow}+1,N_{\downarrow}) 
- \ E_0(N_{\uparrow},N_{\downarrow}) \ ]
\label{chargegap}
\end{equation}
because of the charge-conjugation symmetry. 

Second, a symmetrized DMRG~\cite{ramasesha}
technique is used to calculate
the lowest eigenstates $|n \rangle$ in the $B_u^-$ symmetry sector.
This method yields not only the eigenenergies
$E_n$ of the lowest optically excited states (in particular, the
optical gap $E_{\text{opt}}$), but also
allow us to compute expectation values
$\langle n | \hat{O} | n\rangle$ and thus to analyze the nature
of these states.
To optimize the DMRG program performance, my implementation
of the charge-conjugation and spin-flip symmetries
differs from the original idea presented in 
Ref.~\onlinecite{ramasesha}. 
This is explained in detail in appendix.

Finally, the dynamical DMRG method~\cite{JGE,jeckel}
is used to compute the optical conductivity~(\ref{sigma})
convolved with a Lorentzian distribution of width $\eta > 0$.
Comparisons with exact results
have shown that DDMRG is a very reliable  
numerical method, which yields spectra with an accuracy
comparable to exact diagonalizations but for much larger
systems.~\cite{JGE,EGJ,jeckel}

All DMRG methods have a truncation error which is 
reduced by increasing the number~$m$ of density-matrix 
eigenstates kept 
(for more details, see Refs.~\onlinecite{steve} and 
\onlinecite{dmrgbook}).  
Varying $m$ allows one to compute
physical quantities (including spectra)
for different truncation errors and thus to obtain
error estimates on these quantities.
I have systematically used this procedure to
estimate the precision of my numerical calculations
and adjusted the maximal number $m$ of density-matrix 
states to reach a desired accuracy.
The largest number of density-matrix eigenstates used
in this work is $m=1000$.
For all numerical results presented in this paper 
DMRG truncation errors are negligible.

All numerical calculations have been performed on lattices with
an even number $N$ of sites using open boundary conditions.
As we are interested in the properties of the 
Hamiltonian~(\ref{hamiltonian}) in the thermodynamic limit,
numerical calculations have always been carried out for several system
sizes $N$ in order to investigate finite-size effects. 
The largest system size used here is $N=512$. 
If necessary, the results have been extrapolated to the infinite system 
limit $N \rightarrow \infty$. 
To evaluate finite-size effects in a continuous spectrum
one has to compute it for different sizes while keeping
$\eta N$ = const.~\cite{jeckel}
In this work $\eta N= 12.8 t$ is used.
For all numerical results presented in this paper
finite-size effects (including chain-end effects)
are negligible unless discussed explicitly.
For spectra this means that finite-size effects
are completely hidden by the broadening $\eta$. 
More precisely, DDMRG results for finite $N=12.8t/\eta$ 
are not distinguishable from the corresponding infinite-system spectra
convolved with a Lorentzian distribution of width $\eta$
(see the discussion in Ref.~\onlinecite{jeckel}).

\section{Results \label{sec:results}}

To facilitate the comparison with analytical results, the discussion
of optical excitations in the Mott insulating phase
is divided in three subsections: the limit of a large
Mott gap, the regime of finite Mott gaps, and the
limit of a small Mott gap.
Note, however, that the Mott gap $E_M$ just fixes the energy scale;
the minimal energy required to create a charge excitation is
$E_M/2$ but  
optical excitations do not differ qualitatively as $E_M$ varies
if everything else is kept constant.
In all cases, there is spin-charge separation and the spin sector is 
gapless.
Elementary excitations in the charge sector are spinless bosons in the
lower and upper Hubbard bands.
Optical excitations are always made of an even number ($\geq 2$) of
elementary excitations with opposite charges
(to preserve charge neutrality).
The different types of optical excitations and optical spectra found
in the model~(\ref{hamiltonian}) result from the residual interactions
(essentially the non-local part of the Coulomb repulsion, here $V$) 
between the elementary charge excitations. 

\subsection{Limit of a large Mott gap \label{sec:largegap}}

In the strong-coupling limit $U \gg t$, the properties of the 
model~(\ref{hamiltonian}) in the Mott insulating phase can be described 
using simple concepts.
In the ground state double occupation is prohibited and there is 
exactly one electron on each site.
Elementary charge excitations can be represented as an empty site
(holon in the lower Hubbard band) or a doubly occupied site (doublon in
the upper Hubbard band).
The minimal energy required to create a holon or a doublon is
$E_M/2 = U/2 - O(t) \gg t$.
Optical excitations always consist of an equal number of holons and 
doublons to conserve the total charge.
The ionicity of excited states is defined as the change in the number
of doubly occupied sites with respect to the ground state
\begin{equation}
I_n =  \langle n | \hat{N}_{d} | n \rangle -
\langle 0 | \hat{N}_{d} | 0 \rangle \ ,
\label{ionicity}
\end{equation}
where $\hat{N}_{d} = 
\sum_l \hat{n}_{l,\uparrow}  \hat{n}_{l,\downarrow}$
and $| 0 \rangle, | n \rangle$ denote the ground state and excited 
states, respectively.
Thus, $I_n$ is a measure of the number of doublons (or equivalently of
holons) created by an excitation.
Depending on the strength of the nearest-neighbor interaction parameter
$V$, the low-energy optical excitations are made
of a single doublon-holon pair ($I_n =1$) or are collective 
excitations of several such pairs ($I_n >1$).
Note that $I_n$ is also equal to the derivative of the excitation energy
$E_n - E_0$ with respect to $U$.

\subsubsection*{Single holon-doublon pair}

\begin{figure}
\includegraphics[width=5cm]{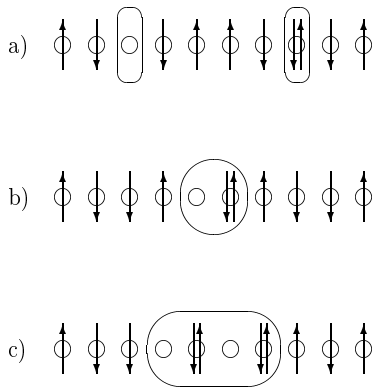}
\caption{\label{fig1}
Schematic representation of (a) an unbound holon-doublon pair,
(b) an exciton, and (c) a biexciton in the strong-coupling limit
$U-2V \gg t$.}
\end{figure}

\begin{figure}
\includegraphics[width=7cm]{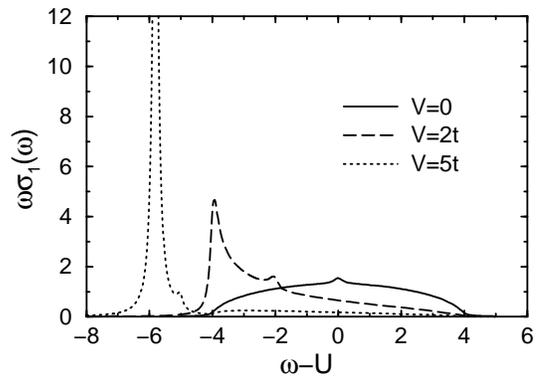}
\caption{\label{fig2}
Reduced optical conductivity $\omega \sigma_1(\omega)$ 
in the limit of a large Mott gap ($U \gg t$)
calculated with DDMRG
for three different values of $V$ using $\eta/t=0.1$ ($N=128$ sites).}
\end{figure}

For $V < U/3 +O(t)$, optical excitations consist of a single
holon-doublon pair and the optical properties, which can be calculated
exactly,~\cite{florian,gallagher} depend only on the parameters 
$V$ and $t$.
For $0 \leq V \leq 2t$, holon and doublon are independent.
A schematic representation of this state is shown in Fig.~\ref{fig1}.
This pair of free charge excitations gives rise to a continuous band in 
the optical spectrum $\sigma_1(\omega)$.
The band starts at the Mott gap $E_M = U -4t$
and has a width of $8t$.
As there is no optical excitation with a lower energy than $E_M$, 
the Mott gap is also the optical gap $E_{\text{opt}}$.
The optical spectra for $V=0$ and $V=2t$ are shown in Fig.~\ref{fig2}
with a broadening $\eta/t=0.1$.
At the conductivity threshold $\sigma_1(\omega)$ vanishes as 
$\sqrt{\omega - E_M}$ for $V < 2t$ but diverges
as $1/\sqrt{\omega - E_M}$ in the special case $V=2t$.
The optical conductivity also has a small peak at $\omega = U-V$ with
1 \% of the spectral weight.~\cite{EGJ} 
This peak is visible inside the band for $V=0$ and $V=2t$ in 
Fig.~\ref{fig2}.
It corresponds to a bound state made of dispersionless charge 
excitations~\cite{florian} and can be seen as a localized exciton
with size $\xi=1$.

For $V > 2t$, there is also a continuous band starting at
$\omega = E_M = U-4t$ due to independent holon-doublon pairs,
but the lowest optical excitation is now an exciton (i.e., a bound
holon-doublon pair) with an energy $\omega_{\text{exc}} =
E_{\text{opt}} = U-V - 4t^2/V$ (Ref. \onlinecite{florian}).
The term $- 4t^2/V$ is the kinetic energy lowering due to the exciton
center-of-mass motion.
Therefore, in the strong-coupling limit ($U \gg t$)
the exciton binding energy is
$E_b = E_M -E_{\text{opt}} = V -4t + 4t^2/V$ ($V > 2t$). 
This binding energy significantly differs from the incorrect
result $E_b=V$ often reported in the literature,~\cite{shuai}
which is (approximately) valid only under the additional condition 
$V \gg t$.  
In the optical spectrum $\sigma_1(\omega)$ 
the exciton generates an isolated $\delta$-peak at 
$\omega_{\text{exc}}$ below the band onset. 
For $V \gg t$, an exciton is essentially the nearest-neighbor
holon-doublon pair shown in Fig.~\ref{fig1}.
This is exactly the state generated by the current 
operator~(\ref{current}) applied to the ground state with one electron
on each site.  
Thus the spectral weight is concentrated in the excitonic peak for
$V \gg t$.
This strong excitonic peak is already clearly visible for $V=5t$ 
in Fig.~\ref{fig2}.
For finite $V/t$, however, there is a finite probability
of finding holon and doublon at a distance $m > 1$.
This probability can be calculated exactly~\cite{EGJ}
\begin{equation}
C(m) = C (1-\delta_{m,0}) e^{-\kappa m}
\end{equation}
with $\kappa = 2 \ln(V/2t)$ and a normalization constant $C$.
The exciton size (the average holon-doublon distance) is then
\begin{equation}
\xi(V > 2t) = \frac{V^2}{V^2-4t^2}
\label{excitonsize}
\end{equation}
and decreases as $V$ increases.  
Correspondingly, one observes a progressive transfer of spectral weight
from the band above $E_M$  to the excitonic peak at 
$\omega_{\text{exc}}=U-V - 4t^2/V$
as $V$ increases.
Note that, as for $V \leq 2t$, there is a small peak corresponding
to a localized exciton at $\omega = U-V$ in the optical conductivity.
This peak lies in the band for $V < 4t$ but is situated
between the band and the strong excitonic peak for $V >4t$.
In Fig.~\ref{fig2} it can be seen for $V=5t$ as a small bump 
at the foot of the strong peak.

Figure~\ref{fig2} shows optical spectra calculated with the DDMRG method
on a 128-site lattice.
On the scale of this figure, the DDMRG spectra are indistinguishable
from the analytical results for an infinite system.~\cite{florian,EGJ}
(For an expanded view showing small deviations, see Fig.~1 of 
Ref.~\onlinecite{EGJ}.)
This perfect agreement confirms the validity of the strong-coupling 
calculations done in Ref.~\onlinecite{florian}.
Moreover, it confirms once more that DDMRG can accurately
reproduce infinite-system optical spectra.~\cite{jeckel}

\subsubsection*{Collective excitations}

As we have just seen, the creation energy of an excited state with 
ionicity
$I_n=1$ is either $U-O(t)$ for an unbound holon-doublon pair or
$U-V-O(t^2/V)$ for an exciton.
Once a first excitation has been created, however, the creation of a 
second holon-doublon pair bound to the first excitation requires only
an energy $U-2V$.
Therefore, when $V$ becomes large enough, the lowest optical excitations
are bound states of $n_{\text{exc}}$ excitons, called excitonic 
strings.~\cite{mazumdar,pleutin}
A biexciton ($n_{\text{exc}}=2$) is shown in Fig.~\ref{fig1} as an 
illustration (see also Ref.~\onlinecite{guo}).

The excitation energy of a $n_{\text{exc}}$-exciton string is
\begin{equation}
E(n_{\text{exc}}) = U -V + (n_{\text{exc}}-1) (U-2V) - O(t^2/V) \ .
\end{equation}
Here the correction of order $t^2/V$ corresponds to the 
kinetic energy lowering due to the center-of-mass motion.
An $n_{\text{exc}}$-exciton string ($n_{\text{exc}} \geq 2$)
appears in the low-energy excitation 
spectrum, around or below the onset $E_M = U -4t$
of the band of free
holons and doublons, if $E(n_{\text{exc}}) \alt E_M$ or
\begin{equation}
V \agt  \frac{n_{\text{exc}}-1}{2 n_{\text{exc}} -1} U + O(t) \ .
\end{equation}
Thus, the biexciton becomes a low-energy excitation for $V \agt U/3$
and longer excitonic strings ($n_{\text{exc}} \geq 3$)
for larger $V$.
The case $n_{\text{exc}} = 1$ corresponds to the usual exciton,
which appears in the low-energy spectrum as soon as
$V > 2t$ as discussed above.
As an $n_{\text{exc}}$-exciton string is made of $n_{\text{exc}}$ 
doublons and holons bound together, it is a neutral excitation,
its ionicity is $I = n_{\text{exc}}$,
and its length is $2n_{\text{exc}}-1$ in units of the lattice constant.
Excitonic strings have been observed in the \textit{non-linear}
optical spectrum of quasi-one-dimensional neutral mixed-stack 
charge-transfer solids and are known to contribute to the 
\textit{non-linear} optical conductivity of models such as the
extended Hubbard model~(\ref{hamiltonian}).\cite{guo,mazumdar}
Naively, one does not expect excitonic strings with 
$I=n_{\text{exc}} \geq 2$ to contribute to the 
\textit{linear} optical spectrum $\sigma_1(\omega)$. 
In the limit $U/t \gg t$
the current operator~(\ref{current}) creates at most one holon-doublon
pair and thus in Eq.~(\ref{currentcorrel})
the matrix elements $\langle n | \hat{J} | 0 \rangle$
between an excited state $|n\rangle$ and the ground state $| 0 \rangle$
must vanish if the ionicity~(\ref{ionicity}) is larger than 1.
Yet, we will see in the next section that excitonic strings with
$n_{\textit{exc}} \geq 2$ are visible in the \textit{linear} optical
conductivity of the extended Hubbard model for large but finite 
couplings $U/t$ and $V/t$.
The reason is that for any finite $t$ there are quantum charge 
fluctuations (virtual holon-doublon pairs) in all eigenstates of
the Hamiltonian~(\ref{hamiltonian}) which leads to small but finite
matrix elements $\langle n | \hat{J} | 0 \rangle$ even if the average
ionicity $I_n$ of an excitation $| n \rangle$ exceeds one,  
\begin{equation}
\langle n | \hat{J} | 0 \rangle \sim \left 
( \frac{t}{U-2V} \right )^{(I_n-1)} \ .
\end{equation}
Thus, for $V \agt U/3 \gg t$ the low-energy optical spectrum
consists of a strong excitonic peak at $\omega = E_{\text{opt}} = 
U-V - O(t^2/V)$ followed by  
several [$n_{\text{exc}} = 2,3,\dots \alt (U-V)/(U-2V)$]
weaker isolated $\delta$-peaks with exponentially decreasing spectral
weight at $\omega \approx E_{\text{opt}} + (n_{\text{exc}}-1)(U-2V)$.
All these peaks appears below (or about) the onset of a weak continuum
due to free holons and doublons at $\omega=E_M$.

As long as $U-2V \gg t$, excitonic strings retain a well-defined
size represented by an integer number $n_{\text{exc}}$ because
the kinetic energy lowering $\sim t$ due to size fluctuations
is much smaller than the energy cost
$\sim (U-2V)$ per exciton in the string.
Close to the phase boundary ($U \approx 2V$) between the CDW ground 
state and the Mott insulator,~\cite{hirsch,jeckel2} however,
size fluctuations become advantageous. 
Thus, for $U-2V \alt t$ low-energy excitations (of the Mott insulator) 
are CDW droplets,
which can be understood as superpositions of excitonic strings of 
every size,
\begin{equation}
| \psi \rangle = c_1 | n_{\text{exc}}= 1 \rangle + 
c_2 | n_{\text{exc}} =2 \rangle + 
c_3 | n_{\text{exc}} =3 \rangle + \dots
\label{droplet}
\end{equation}
with a broad distribution of coefficients $c_n$ ($\sum_n |c_n|^2 =1$).
For $U-2V \rightarrow 0^+$ the distribution becomes flat,
i.e., $|c_n|^2 \rightarrow$ const.
(For comparison, an $n_{\text{exc}}$-exciton string can be described
by the above state with
$|c_n|^2 \approx 1$ for $n=n_{\text{exc}}$ and $|c_n|^2 \ll 1$
for $n \neq n_{\text{exc}}$.)
CDW droplets in the Mott insulating phase are the analogue to the SDW 
droplets in the CDW insulator discussed by Hirsch.~\cite{hirsch}
As excitonic strings, these CDW droplets are neutral excitations,
but one can generalize the concept to CDW droplets carrying charges 
(see below).
The average size $r_{\text{CDW}}$ of a CDW droplet is related to its
ionicity by $r_{\text{CDW}} = 2 I = 2 \sum_i i |c_i|^2$. 
Its excitation energy is
\begin{equation}
E(r_{\text{CDW}}) = U- V + \frac{r_{\text{CDW}}}{2} (U-2V) -s t \ ,
\end{equation}
where  $s >0$ and $-st$ represents the kinetic energy 
lowering due to droplet size fluctuations and center-of-mass motion
($s \approx 4$ for $U-2V \rightarrow 0^+$).
Contrary to excitonic strings, the ionicity $I$ of a CDW droplet is
not a integer number but can take any value $\geq 1$.
Therefore, for $U-2V \ll t$ there is a band of CDW droplet
excitations starting at $U-V -st$.
Moreover, the matrix element $\langle n | \hat{J} | 0 \rangle$
for a CDW droplet $| n \rangle$ is essentially given by the
overlap $c_1$ with the single exciton state in Eq.~(\ref{droplet}).
Thus in this regime one expects that the CDW droplets give rise to a 
band in the optical spectrum $\sigma_1(\omega)$ starting
at $\omega = E_{\text{opt}} = U-V -st$.
This band lies below the Mott gap $E_M$.
It should be noted  that the Mott gap is determined by the excitation
energy of unbound holons and doublons, $E_M \approx U -4t$, as long
as $U-2V \gg t$. 
For $U-2V \alt t$, however, CDW droplets carrying a charge $\pm e$
[a CDW droplet~(\ref{droplet}) bound to an extra holon or doublon]
have a lower energy than a bare holon or doublon 
and reduce the gap for charge excitations~(\ref{chargegap}) to
$E_M = U -s't$ with $s' \approx 8$ for $V \rightarrow U/2$.

In summary, in the limit of a large Mott gap ($U \gg t$) 
there are four distinct regimes corresponding to four types of
excitations in the low-energy optical spectrum:
independent charge excitations for $V \leq 2t$, 
excitons for $V > 2t$ but $V \alt U/3 $, excitonic strings
for $V \agt U/3 \gg t$ but $U-2V \agt t$,
and CDW droplets for $U-2V \alt t$.
We will see that these excitations are also found in the
optical spectrum of the extended Hubbard model for finite
interaction strengths and Mott gaps.

\subsection{Regime of finite Mott gaps \label{sec:intermediate}}

\begin{figure}
\includegraphics[width=7cm]{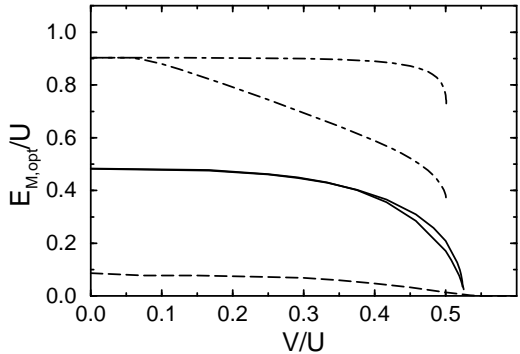}
\caption{\label{fig3}
Mott gap $E_M$ (upper line) and optical gap $E_{\text{opt}}$ 
(lower line) versus $V$
for $U/t=40$ (dot-dashed), $6$ (solid), and $2$ (dashed).
For $U=2t$, $E_{\text{opt}} = E_M$.
}
\end{figure}

For finite coupling parameters $U$ and $V$ the low-energy optical 
properties of the extended Hubbard model~(\ref{hamiltonian}) can be 
calculated using the ground-state and symmetrized DMRG methods
presented in Sec.~\ref{sec:methods}.
Figure~\ref{fig3} shows the Mott gap $E_M$ 
and the optical gap $E_{\text{opt}}$ as a function
of $V$ for three values of $U$.
Both gaps increase monotonically with $U$ but decrease with increasing
nearest-neighbor interaction $V$.~\cite{shuai,kancharla}
For all values of $U$ the optical gap equals the Mott gap
(in the thermodynamic limit) as long as $V \leq 2t$ but for larger $V$,
$E_{\text{opt}}$ becomes smaller than $E_M$.
This suggests that for all $U>0$ the low-energy excitations 
are unbound for $V \leq 2t$ and bound for $V > 2t$ as in the
$U \gg t$ limit.~\cite{florian,shuai}
Obviously, the condition $V > 2t$ can be realized only for relatively 
strong coupling ($U \agt 4t$) because the Mott insulating phase
exists only for $V$ up to $V_c \approx U/2$.
The Mott gaps in Fig.~\ref{fig3} are initially 
almost constant as $V$ 
increases then diminish significantly close to the phase boundary
$V_c$.
This agrees with the strong-coupling analysis in the previous section,
which suggests that $E_M$ is essentially independent of $V$
for $V_c-V \gg t$ but is reduced by quantity $\propto t$
as $V$ approaches the critical value $V_c$.  
Note that on the critical line between the CDW and Mott insulating 
phases, both gaps vanish for $U \leq 3t$ 
while the Mott gap clearly remains finite for stronger coupling
($U \geq 4t$).

\begin{figure}
\includegraphics[width=6cm]{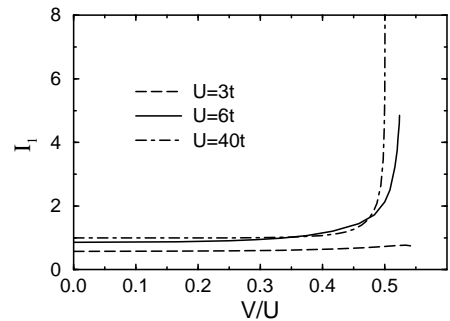}
\caption{\label{fig4}
Ionicity $I_1$ of the first optically excited state ($1B_u^-$) 
as a function of $V$ for three different values of $U$.
}
\end{figure}

To determine the nature of the low-energy optical excitations
I have calculated their ionicity~(\ref{ionicity}).
Figure~\ref{fig4} shows the ionicity $I_1$ of the first
optically excited state (the $1B_u^-$ state) as a function of $V$
for three values of $U$.
In the half-filled Hubbard model ($V=0$) $I_1$ increases monotonically
from 0 at $U=0$ to 1 for $U \rightarrow \infty$, reflecting the 
increasingly ionic nature of the elementary charge excitations
in the lower and upper Hubbard bands.
The ionicity increases slowly with $V$ and remains below or close to 1
for most couplings ($U,V$), which confirms that the corresponding
optical excitations are made of a single pair of elementary
charge excitations.
In the regime $U \approx 2V > 4t$, however, one observes a rapid
but continuous 
increase of $I_1$ to values larger than 2 as 
$V \rightarrow V_c \approx U/2$. 
This shows that the lowest optical excitation has become a CDW droplet.
Looking at higher optical excitations, one finds the same qualitative
behavior of the ionicity as a function of $U$ and $V$.
Additionally, one observes the formation of excitonic string with 
integer $I = n_{\text{exc}} \geq 2$.

\begin{figure}[b]
\includegraphics[width=6cm]{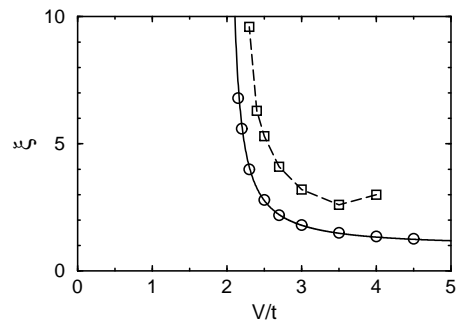}
\caption{\label{fig5}
Exciton size 
as a function of $V$ for $U=40t$ (circles) and
$U=8t$ (squares).
The solid line is the $U \gg t$ result, 
Eq.~(\ref{excitonsize}).
The dashed line is just a guide for the eyes.
}
\end{figure}

To determine whether the pair of elementary charge excitations
form bound (exciton) or unbound states, one can calculate the average
distance between both excitations using an exciton correlation
function.~\cite{joerg,joerg2,EGJ}
For $V < 2t$ I have found that this average distance always diverges 
with increasing system size $N$. 
This result definitively confirms that in this regime 
an optical excitation is a pair of independent charge excitations,
in agreement with the strong-coupling analysis.
For $V > 2t$, the average distance tends to a finite value for
$N \rightarrow \infty$ as expected for an exciton.
The exciton size $\xi$ determined with this procedure is
shown in Fig.~\ref{fig5} as a function of $V$ for two finite values of 
$U$. 
The exciton size in the limit $U \gg t$, Eq.~(\ref{excitonsize}), 
is also plotted for comparison.
The size $\xi$ increases and diverges as $V$ tends to $2t$,
showing the unbinding of the exciton at $V=2t$.
Note that for $U=40t$ the sizes measured with the exciton 
correlation function~\cite{EGJ,joerg,joerg2} agree perfectly with 
Eq.~(\ref{excitonsize}).

One can gain some knowledge about the nature of optically excited
states by looking at the scaling of their spectral weight
\begin{equation}
W_n = \frac{\pi}{Na} \frac{|\langle n | \hat{J} | 0 \rangle|^2}{E_n-E_0}
\label{weights}
\end{equation}
[see Eqs.~(\ref{sigma}) and (\ref{currentcorrel})]
with the system size $N$.
For $V \leq 2t$ I have found that the spectral weight of 
low-lying excitations vanishes for $N \rightarrow \infty$,
which is expected for states belonging to a continuum.
(As there is an infinite number of states in a continuous band, the
spectral weight of each state must go to zero as $N \rightarrow \infty$,
so that the total weight in any finite frequency interval remains
finite.)
For $V > 2t$ (but outside the CDW droplet regime), I have found that
the optical weight $W_1$ of the $1B_u^-$ state tends to a finite value 
in the thermodynamic limit.
This corresponds to a $\delta$-peak (with total weight $\geq W_1$)
in the optical spectrum as expected for an exciton or an
excitonic string.
In the CDW droplet regime, finite-size effects become large and complex
and, in most cases, it has not been possible to determine the scaling of
the spectral weights~(\ref{weights}).

\subsubsection*{Optical spectra}

The above analysis shows that low-energy optical excitations
in the regime of finite Mott gaps are identical to those
found in the limit of a large Mott gap and can
be interpreted using the simple theory developed
for the strong-coupling limit in Sec.~\ref{sec:largegap}.  
Turning next to the optical spectrum I have calculated 
$\sigma_1(\omega)$ for various parameters 
$40t \geq U \geq 3t$ and $U/2 \agt V \geq 0$ using the DDMRG method.
I have found that the optical spectra of systems with a finite Mott gap
closely resemble those observed in the limit of a
large Mott gap.  

\begin{figure}
\includegraphics[width=7cm]{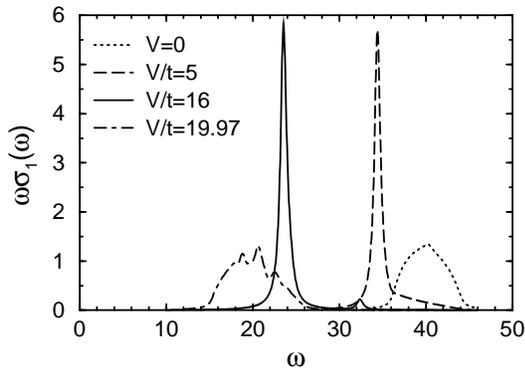}
\caption{\label{fig6}
Reduced optical conductivity $\omega \sigma_1(\omega)$
for $U=40t$ and four different values of $V$
calculated using DDMRG with $\eta/t=0.4$ ($N=32$ sites). 
}
\end{figure}

As a first example, the optical conductivity $\sigma_1(\omega)$ 
is shown in Fig.~\ref{fig6} for $U=40t$ and several values of $V$
representing the four different regimes:
free charge excitations ($V=0$), excitons ($V=5t$), excitonic strings
($V=16t$), and CDW droplets ($V=19.97t$).
For $V=0$ there is a single continuous band starting at
$\omega=E_{\text{opt}}=E_M= 36.14t$. 
At the band edges the optical conductivity vanishes as 
$\sqrt{\omega-E_M}$ as discussed in Ref.~\onlinecite{JGE}.

For $V=5t$ a strong excitonic $\delta$-peak appears at
$\omega_{\text{exc}} = E_{\text{opt}} = 34.39t$  below 
the Mott gap $E_M = 36.13t$.
The exciton has a size $\xi \approx 1.2$ in perfect agreement with
Eq.~(\ref{excitonsize}).
There is also a weak continuous band of free charge excitations above 
$E_M$, which is only visible as a high-frequency tail of the exciton 
peak in Fig.~\ref{fig6}.
The gap between the excitonic peak and the band is not visible in
Fig.~\ref{fig6} because of the large broadening $\eta/t=0.4$ used here,
but it can be checked with a scaling analysis for $\eta \rightarrow 0$
($N \rightarrow \infty$) as discussed in Ref.~\onlinecite{jeckel}.
The only qualitative difference between the present result for 
$U=40t$
and the corresponding result in the limit $U \gg t$ 
(see Fig.~1 in Ref.~\onlinecite{EGJ})
is the absence of 
the weak peak associated with a localized exciton at $\omega=U-V$.
Nevertheless, this weak peak is not an artifact of the
strong-coupling limit because its existence has been
confirmed in the Hubbard model ($V=0$) down to $U=4t$ 
(Ref.~\onlinecite{JGE}).
The finite spectral weight carried by the localized exciton originates
from a ground-state dimer-dimer correlation of the spin degrees of 
freedom.~\cite{florian}
In the strong-coupling limit ($U-2V \gg t$) of the extended Hubbard 
model~(\ref{hamiltonian}), the effective exchange coupling between
nearest-neighbor spins depends on the occupation of the neighboring
sites if $V\neq 0$. Thus the effective spin Hamiltonian is not
the one-dimensional Heisenberg model, in general. 
Only for $V=0$ or in the limit $U \gg V$, 
the effective spin Hamiltonian reduces to the Heisenberg model
with a constant exchange coupling $J=4t^2/U$. 
In this case, the ground state has the relevant 
spin dimer-dimer correlations 
and the localized exciton carried a finite optical weight
as explained in detail in Ref.~\onlinecite{florian}.
For finite $U$ and $V$, however, the spin dimer-dimer correlation
is presumably destroyed by the fluctuations of the spin exchange 
coupling 
and thus the optical weight of the localized exciton vanishes.

\begin{figure}
\includegraphics[width=6cm]{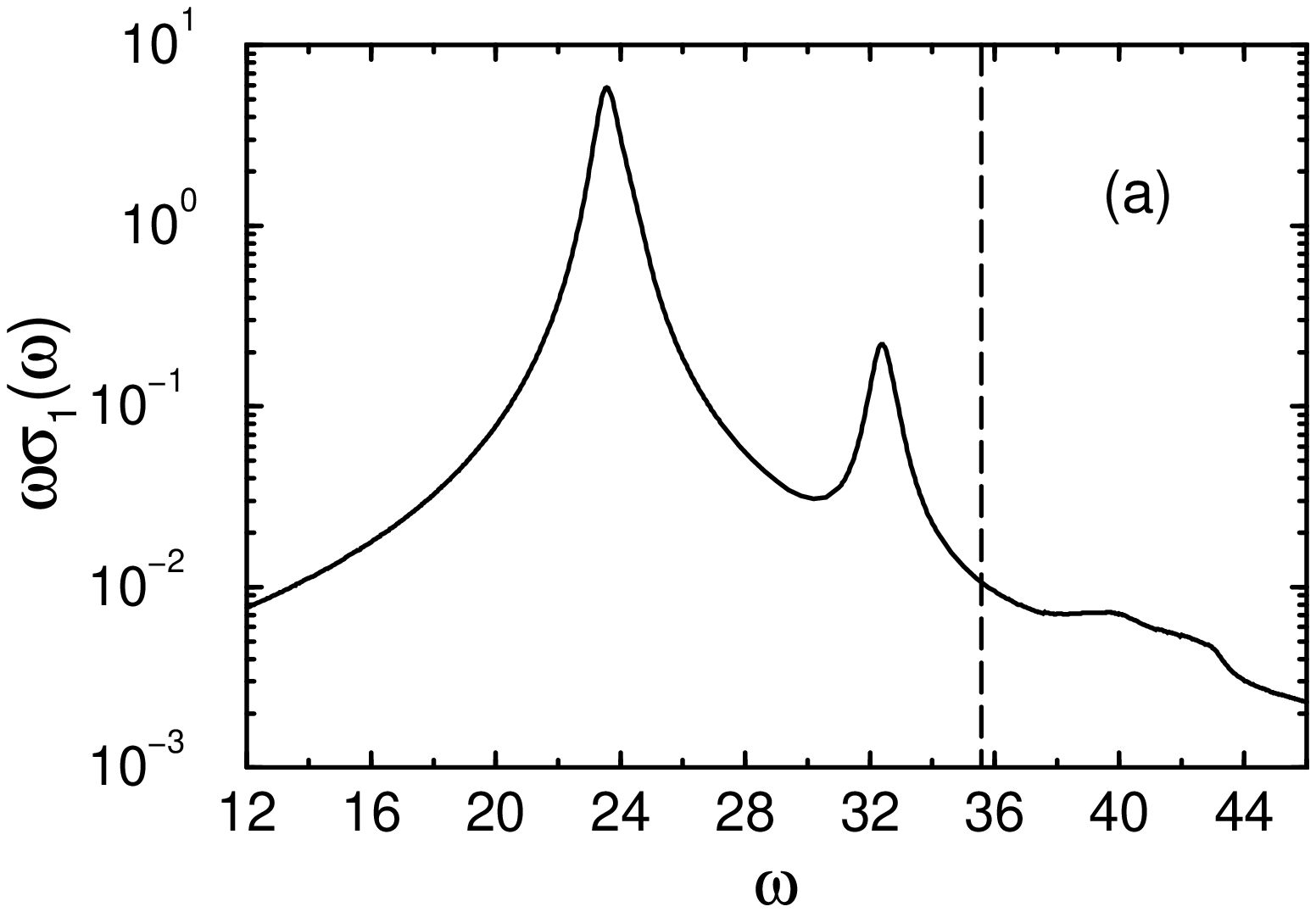}
\includegraphics[width=6cm]{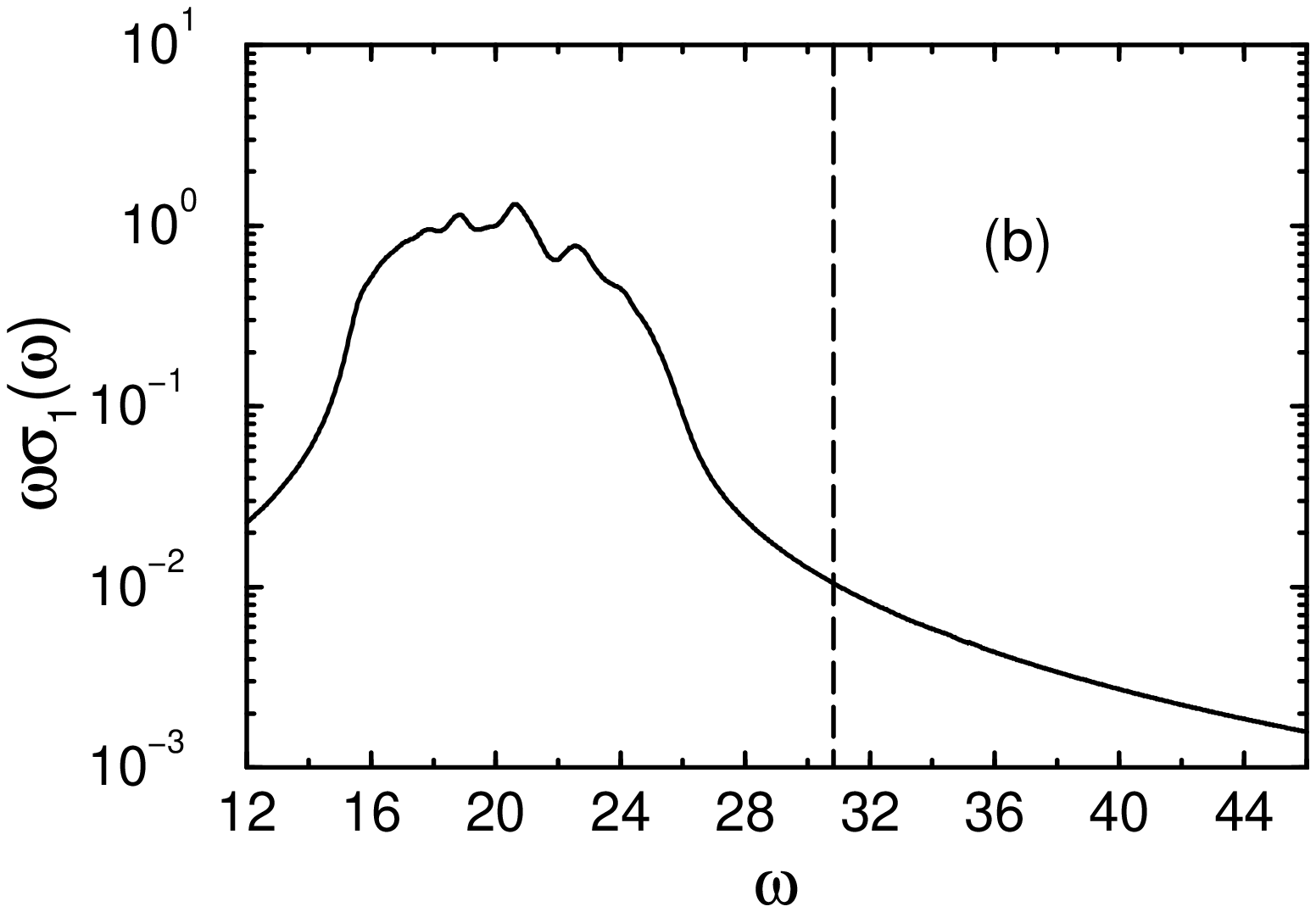}
\caption{\label{fig7}
Reduced optical conductivity $\omega \sigma_1(\omega)$
for $U=40t$ and (a) $V=16t$ and (b) $V=19.97t$   
calculated using DDMRG with $\eta/t=0.4$ ($N=32$ sites). 
Vertical lines indicate the Mott gap $E_M$.
}
\end{figure}

For $V=16t$ the condition $V \agt U/3$ is satisfied 
and excitonic strings appear in the optical spectrum below or around 
the Mott gap $E_M = 35.58t$.
As seen in Fig.~\ref{fig6},
most of the spectral weight is concentrated in the exciton
of size $\xi=1.0$
at $\omega_{\text{exc}} = E_{\text{opt}} = 23.53t \approx U-V$.
The biexciton at $\omega = 32.35t 
\approx 2U-3V$ is barely visible in Fig.~\ref{fig6}.
The optical conductivity $\sigma_1(\omega)$ is again shown
in Fig.~\ref{fig7}(a) on a logarithmic scale.
The isolated peaks associated with both excitations are now
clearly visible.
The measured ionicity~(\ref{ionicity}) is $I=1.1$ and $I=2.2$ for
the exciton and the biexciton, respectively. 
In Fig.~\ref{fig7}(a) the remnant of the continuous band of free charge
excitations and the triexciton (at $\omega \approx 3U-5V = 40t$)
are also visible in the interval $\omega =36-44 t$ above the Mott gap.

For $V=19.97t \approx U/2$ the optical conductivity spectrum
is radically different. 
The excitonic strings collapse into a band of CDW 
droplets with varying sizes.
For instance, the $1B_u^-$ state is a droplet of size
$r_{\text{CDW}} = 8.9$ with an energy $E_{\text{opt}} \approx 15.5t$.
These CDW droplets give rise to a broad band in the optical 
conductivity spectrum shown in Fig.~\ref{fig6}.
The onset of this band is well below the Mott gap $E_M =30.83t$.
On the logarithmic scale of Fig.~\ref{fig7}(b),
one sees that, in this particular case, the entire optical weight 
seems to be below $E_M$ (for $\eta \rightarrow 0$).
The appearance of a band below the Mott gap is also visible in the
current-current correlations for $U=12t$ and $V=6t$
presented in Ref.~\onlinecite{kancharla} 
but the optical spectrum in the regime $U\approx 2V$ 
is not interpreted correctly in that work.

\begin{figure}
\includegraphics[width=7cm]{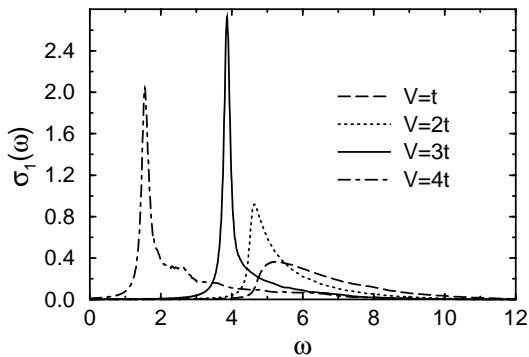}
\caption{\label{fig8}
Optical conductivity $\sigma_1(\omega)$
for $U=8t$ and four different values of $V$
calculated with DDMRG using $\eta/t=0.1$ ($N=128$ sites). 
}
\end{figure}

As a second example and to illustrate the finite-size-scaling analysis 
I have carried out for dynamical spectra,
I discuss the optical conductivity $\sigma_1(\omega)$ for $U=8t$.
Figure~\ref{fig8} shows the evolution of the optical conductivity
for increasing nearest-neighbor repulsion $V$.
For $V=t$ and $V=2t$, the spectrum contains a single continuous
band due to free charge excitations starting at $E_{\text{opt}} 
=E_M = 4.67t$ and $4.53t$, respectively.
For $V=3t$ the spectrum consists of a strong peak 
corresponding to an exciton of size $\xi = 3.2$ and 
energy $\omega_{\text{exc}} = E_{\text{opt}} = 3.86t$, 
and of a weak band above the Mott gap $E_M = 4.10t$.
This band is visible in Fig.~\ref{fig8} as the high-frequency tail of 
the excitonic peak.
For $V=4t$, CDW droplets of varying 
sizes dominate the optical spectrum. 
For instance, the $1B_u^-$ state is a droplet of size 
$r_{\text{CDW}} = 5.6$ with an excitation energy 
$E_{\text{opt}} = 1.55t$
lower than the Mott gap $E_M =2.29t$. 
There is no intermediate regime with well-defined excitonic strings for
this value of $U$.

The precise shape of $\sigma_1(\omega)$ cannot be determined 
from the sole results shown in Fig.~\ref{fig8} because of
the finite resolution and system size used, $\eta/t = 12.8/N = 0.1$.
To determine the properties of $\sigma_1(\omega)$ 
with maximal resolution ($\eta \rightarrow 0$)
in the thermodynamic limit ($N \rightarrow \infty$),
one can perform a scaling analysis with $\eta N = const.$
as explained in Ref.~\onlinecite{jeckel}.
(Here I have used $\eta N = 12.8t$.)
The scaling analysis of
the optical conductivity $\sigma_1(\omega)$ calculated with DDMRG
always yields results which are qualitatively and quantitatively
consistent with the properties of low-lying optical excitations
determined using the ground-state and symmetrized DMRG methods.
For instance, $\sigma_1(\omega)$ vanishes for 
all $\omega < E_{\text{opt}}$ and there is a continuous band 
for $\omega \geq E_{\text{opt}}$ or a $\delta$-peak at 
$\omega = E_{\text{opt}}$ in the limit 
$\eta \sim 1/N \rightarrow 0$.

\begin{figure}
\includegraphics[width=6cm]{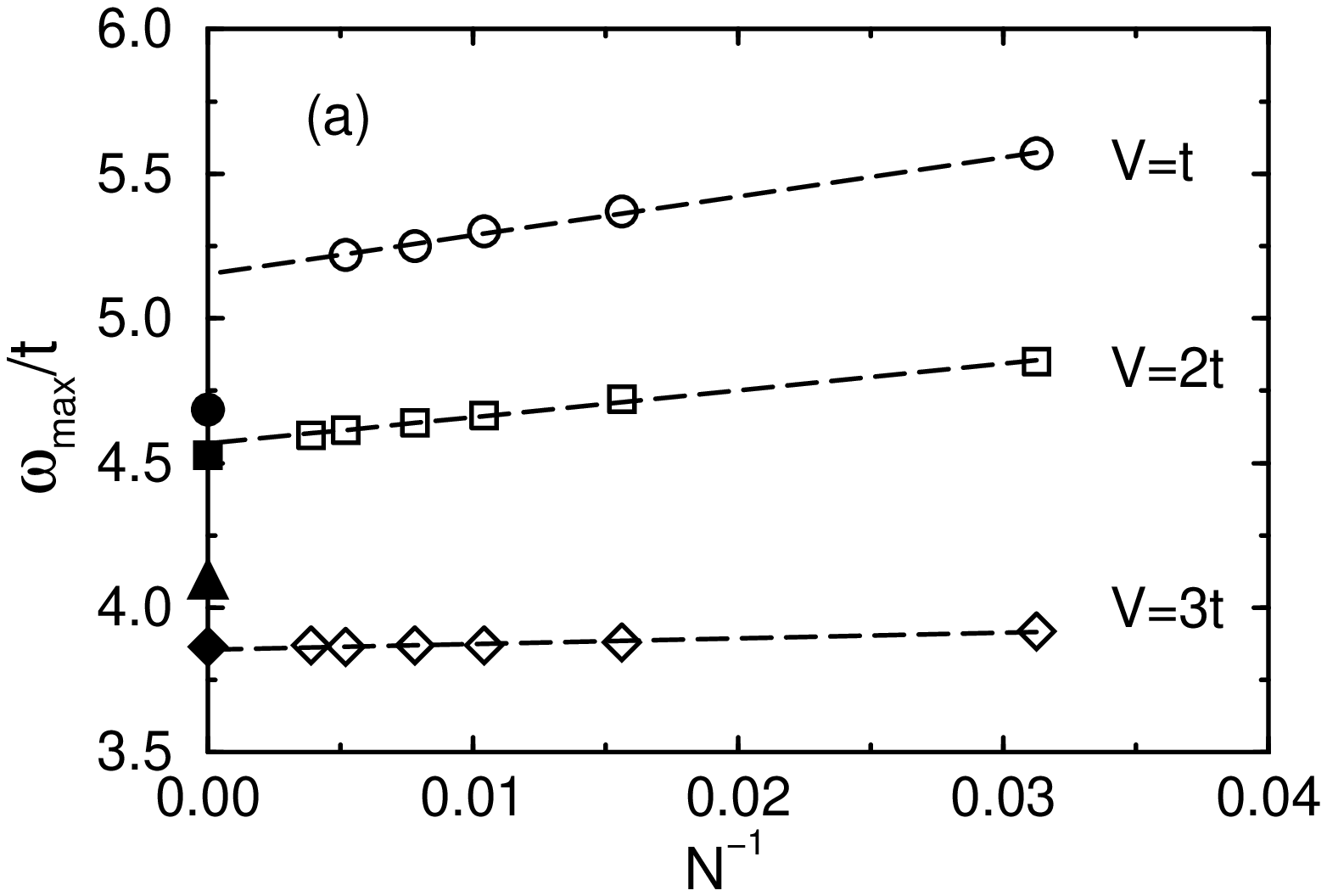}
\includegraphics[width=6cm]{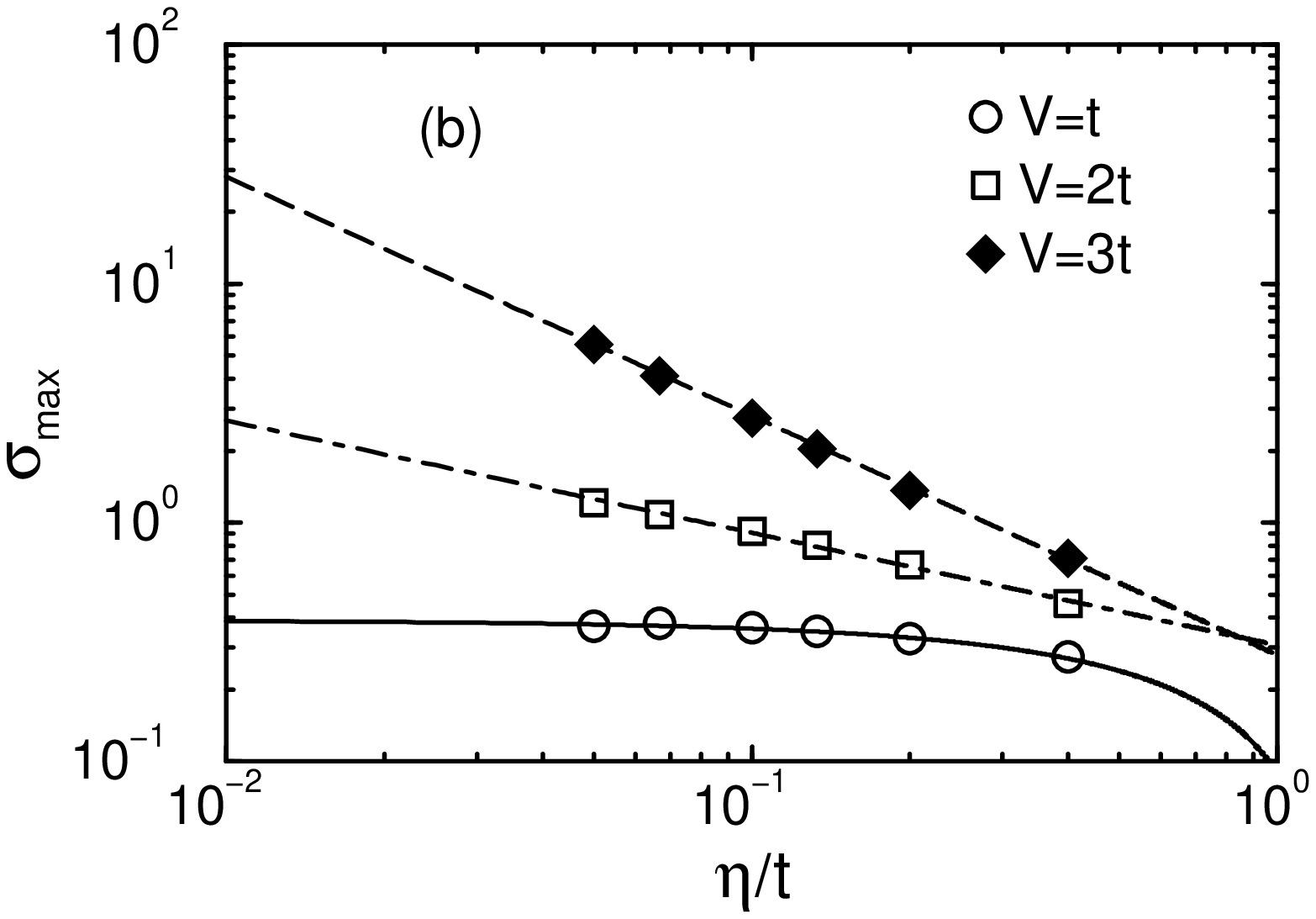}
\caption{\label{fig9}
Scaling analysis of the maximum in DDMRG optical spectra 
$\sigma_1(\omega)$ for $U=8t$ and three different 
values of $V$ ($\eta N =12.8t$).  
(a) Position $\omega_{\text{max}}$ of the maximum (open symbols)
as a function of the inverse system size.
Lines are linear fits to these data.
Solid symbols show the optical gaps $E_{\text{opt}}$
calculated with symmetrized DMRG for
$N \rightarrow \infty$.
For $V=3t$, the Mott gap $E_{M}$ is indicated by a triangle.
For $V/t=1$ and $2$, $E_{M} =E_{\text{opt}}$.
(b) Maximum of $\sigma_1(\omega)$ as a function of $\eta/t$.
Lines are fits to the numerical data: the dashed line corresponds
to $\eta^{-1}$, the dot-dashed to $\eta^{-1/2}$, and the solid line is 
a linear fit in $\eta$.  
}
\end{figure}

The scaling analysis of
the conductivity maximum $\sigma_{\text{max}} =
\sigma_1(\omega_{\text{max}})$ in DDMRG spectra
is illustrated in Fig.~\ref{fig9}
for the
same interaction parameters as in Fig.~\ref{fig8}.
For $V=t$, $\omega_{\text{max}}$ tends 
to a value $(5.15t)$ larger than the optical gap $E_{\text{opt}}=4.67t$
calculated with symmetrized DMRG for $N \rightarrow \infty$
[see Fig.~\ref{fig9}(a)]
while $\sigma_{\text{max}}$ tends to a finite value for 
$\eta \rightarrow 0$ [see Fig.~\ref{fig9}(b)].
Moreover, the derivative of $\sigma_1(\omega)$ has a maximum that 
diverges as $1/\sqrt{\eta}$ for $\eta \rightarrow 0$
at $\omega = E_{\text{opt}}$.
These features correspond to a continuum that vanishes
as $\sqrt{\omega -E_{\text{opt}}}$ at the conductivity
threshold and goes
through a maximum just above the optical gap at $\omega_{\text{max}}
\approx 1.1 E_{\text{opt}}$ (see Ref.~\onlinecite{jeckel}).
For $V=2t$, $\omega_{\text{max}}$ tends to the same value as 
the optical gap $E_{\text{opt}}$ for $N \rightarrow \infty$
[see Fig.~\ref{fig9}(a)]
and $\sigma_{\text{max}}$ diverges as $1/\sqrt{\eta}$ for
$\eta \rightarrow 0$ [see Fig.~\ref{fig9}(b)].
These features correspond to a continuum that diverges
as $1/\sqrt{\omega -E_{\text{opt}}}$ at the conductivity
threshold.~\cite{jeckel}
(Note that for all values of $U$ investigated
$\sigma_1(\omega)$ displays this divergence at $V=2t$.) 
Therefore, the features of the optical spectrum for $V \leq 2t$  
are similar to those found in the strong-coupling limit.
For $V=3t$, $\omega_{\text{max}}$ tends for $N \rightarrow \infty$
to the same value as the optical gap $E_{\text{opt}}$ (which is smaller
than the Mott gap in the thermodynamic limit) [see Fig.~\ref{fig9}(a)]
and $\sigma_{\text{max}}$ diverges as $1/\eta$ for $\eta \rightarrow 0$
[see Fig.~\ref{fig9}(b)].
These features correspond to a $\delta$-peak at 
$\omega = E_{\text{opt}}$.
Moreover, $\sigma_1(\omega)$ vanishes between $E_{\text{opt}}$ and
$E_M$ but remains finite above $E_M$ 
in the limit $\eta \sim 1/N \rightarrow 0$.
Therefore, in the thermodynamic limit the spectrum for $V=3t$
(shown in Fig.~\ref{fig8} for $N=128$ sites)
consists of an excitonic $\delta$-peak 
separated from the band of independent charge excitations
as in the strong-coupling limit.
In the CDW droplet regime (i.e., close to the
critical line $V_c \approx U/2$ separating Mott and CDW phases)
finite-size effects are more complicated and larger than in the
other regimes.
As a consequence, for $V=4t$ it has not been possible to perform a 
conclusive analysis with the largest system sizes ($N=256$) available. 
It seems that the low-energy spectrum contains a $\delta$-peak
at $E_{\text{opt}}$ and a band starting immediately above
$E_{\text{opt}}$, both due to CDW droplets.

In summary, I have found that the optical properties for all finite
Mott gaps (i.e., for all $U>0$, $V \geq0$ in the Mott phase)
are qualitatively similar to those calculated
in the limit of a large Mott gap 
(Sec.~\ref{sec:largegap}, Refs.~\onlinecite{EGJ},
\onlinecite{florian}):

(i) For $V \leq 2t$ (that is the only possible case for $U \alt 4t$),
independent charge excitations give rise to a continuous band starting 
at the Mott gap $E_M$, which is equal to the optical gap.
The band width is typically $\sim 8t$.
For $V < 2t$,  $\sigma_1(\omega)$ vanishes smoothly at the 
threshold $E_{\text{opt}}$, typically as $\sqrt{\omega-E_{\text{opt}}}$.
At $V=2t$, $\sigma_1(\omega)$ diverges as 
$1/\sqrt{\omega-E_{\text{opt}}}$
for $\omega-E_{\text{opt}} \rightarrow 0^+$.

(ii) For $V > 2t$ but $U > 2V+O(t)$ and $V < U/3 + O(t)$
(this is possible only for $U \agt 4t$), the optical spectrum 
consists of an excitonic $\delta$-peak below the Mott gap and a band
due to free charge excitations above $E_M$. 
Most of the optical weight is in the excitonic peak for    
$V \agt 3t$.

(iii) If $U$ is large enough ($U \agt 12t$) excitonic strings
appear in the low-energy spectrum for $V > U/3 +O(t)$
but $U > 2V+O(t)$. 
They generate isolated $\delta$-peaks below the Mott gap $E_M$ in 
the optical conductivity $\sigma_1(\omega)$ with
a separation between peaks of $\Delta\omega \approx U-2V$.
The first peak is an exciton and contains most of the spectral weight.
A very weak band due to free charge excitations still exists
above $E_M$.

(iv) Close to the boundary $V_c \approx U/2$ between the Mott and CDW 
phases, if $V$ exceeds $2t$, CDW droplets dominate the low-energy 
spectrum
and give rise to a broad band (including sharp peaks) starting below 
the gap for charge excitations~(\ref{chargegap}).

\subsection{Limit of a small Mott gap \label{sec:smallgap}}

In the limit of a small Mott gap ($E_M \ll t$)
the coherence length $\sim 4t/E_M$
becomes very large and it is not possible to carry out
numerical simulations on lattices large enough ($N \gg 4t/E_M$)
to determine the optical spectrum with confidence.
Fortunately, in this limit 
field-theoretical methods provide generic results
for the low-energy optical spectrum of a one-dimensional
Mott insulator.~\cite{JGE,controzzi,EGJ}
Field-theoretical results
are applicable to lattice models such as Eq.~(\ref{hamiltonian})
for gaps up to $E_M \alt t$,
which makes possible a direct quantitative comparison of field theory
and DDMRG calculations.~\cite{JGE,EGJ,jeckel}

\begin{table}
\caption{ \label{table1}
Mott gap $E_M$, optical gap $E_{\text{opt}}$, ionicity $I_1$ of the
first optically excited state $1B_u^-$, and the corresponding 
field-theory interaction
parameter $\beta^2$ (see text) 
for several values of $U$ and $V$ in the small
gap regime $E_M/t =  0.6-0.7$.
}
\begin{ruledtabular}
\begin{tabular}{cccccc}
$U/t$ & $V/t$ & $E_{M}/t$ & $E_{\text{opt}}/t$ & $I_1$ & $\beta^2$ \\
\colrule    
3& 0  & 0.631 & 0.628 & 0.574 & 1  \\
3.5& 1.4  & 0.664 & 0.662& 0.784 & 0.61  \\
4& 1.9  & 0.628 & 0.627& 1.11 & 0.52 \\
4.15& 2 & 0.645 & 0.642 & 1.20 & 1/2\\
4.5& 2.25  & 0.638 & 0.611 & 1.54 & (0.449) \\
5& 2.57 & 0.605 & 0.524 & 2.22& (0.400) \\
6& 3.115 & 0.643 & 0.445& 4.00 & (0.327) \\
8& 4.137 & 0.641 & 0.24 & 19.9& - \\
\end{tabular}
\end{ruledtabular}
\end{table}

In the field-theoretical approach, elementary charge excitations
are holons (in the lower Hubbard band) and anti-holons (in the
upper Hubbard band). 
Optical excitations are made of a equal number of holons and 
anti-holons.
Assuming that the low-energy excitations consist of one holon-antiholon
pair, the optical conductivity is
\begin{equation}
\sigma^{\text{FT}}_1(\omega) = A \ S_{\beta}(\omega/E_M) \,
\end{equation}
where $S_{\beta}(x)$ is a known function depending on the field-theory
interaction parameter $0 < \beta \leq 1$, and $A$ is a unknown
constant which sets the conductivity scale.
Strictly speaking, this result is exact only for $\omega < 2 E_M$
and $\beta^2 > 1/3$
but it has been found by comparison with DMRG results
that corrections for $\omega > 2 E_M$
are usually negligible.~\cite{JGE,EGJ}
For $\beta^2 \geq 1/2$, $\sigma^{\text{FT}}_1(\omega)$ 
describes a single 
continuous band starting at $E_{\text{opt}} = E_M$, which is
due to independent holons and anti-holons.
The optical conductivity vanishes smoothly as 
$\sqrt{\omega-E_{\text{opt}}}$ for $\beta^2 > 1/2$ and diverges
as $1/\sqrt{\omega-E_{\text{opt}}}$ for $\beta^2=1/2$
at the conductivity threshold.
For $1/3 < \beta ^2 < 1/2$, there is a $\delta$-peak at
$\omega= E_{\text{opt}} < E_M$ in addition of the band starting
at $\omega =E_M$.
The $\delta$-peak is due to a bound holon-antiholon pair (exciton).
For $\beta^2 < 1/3$ additional excitons and excitonic strings
(made of several holon-antiholon pairs) appear in the spectrum.
Therefore, field-theoretical predictions for the optical
conductivity of a one-dimensional Mott 
insulator are qualitatively similar 
to what we have found in the extended Hubbard model~(\ref{hamiltonian})
using a strong-coupling analysis and DDMRG simulations.

\begin{figure}
\includegraphics[width=7cm]{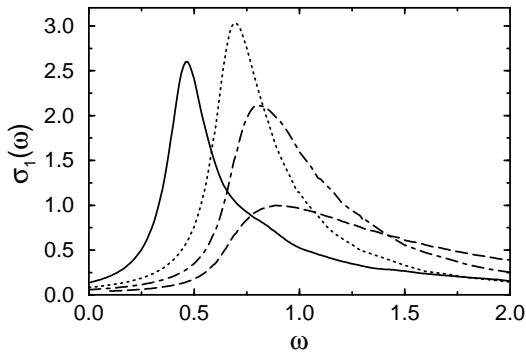}
\caption{\label{fig10}
Optical conductivity $\sigma_1(\omega)$
calculated with DDMRG using $\eta/t=0.1$ ($N=128$ sites)
for $U=3t, V=0$ (dashed), $U=3.5t, V=1.4t$ (dot-dashed), 
$U=4.15t, V=2t$ (dotted),
and $U=6t, V=3.155t$ (solid). 
}
\end{figure}

The field-theory parameters $E_M$, $\beta$, and $A$ 
must be estimated numerically by comparison with DMRG results
because one does not know their relations to the lattice model
parameters $U, V, t$. 
I have first determined several couplings $(U,V)$ which yield
approximately the same Mott gap $E_M/t \approx 0.6-0.7$.
These couplings are listed in Table~\ref{table1} with the corresponding
Mott and optical gaps calculated using the ground-state and symmetrized
DMRG methods.
Then I have calculated the optical conductivity $\sigma_1(\omega)$
for these parameters using DDMRG.
Some results are shown in Fig.~\ref{fig10}.
Note the progressive displacement of spectral weight
to lower energy as $V$ increases although the Mott gap remains
almost constant (see Table~\ref{table1}).
To determine the parameters $\beta$ and $A$ one can now compare the
field-theoretical spectra with the DDMRG data.
[$\sigma_1^{\text{FT}}(\omega)$ has to be convolved with a Lorentzian
distribution of appropriate width $\eta$ to make a direct 
comparison.~\cite{jeckel}]
This procedure yields the parameter $\beta^2$ listed in 
Table~\ref{table1}.
As expected the boundaries between independent charge excitations and 
excitons in the field theory and in the lattice 
model~(\ref{hamiltonian}) coincide: 
$\beta^2 = 1/2$ corresponds to $V=2t$.

\begin{figure}
\includegraphics[width=7cm]{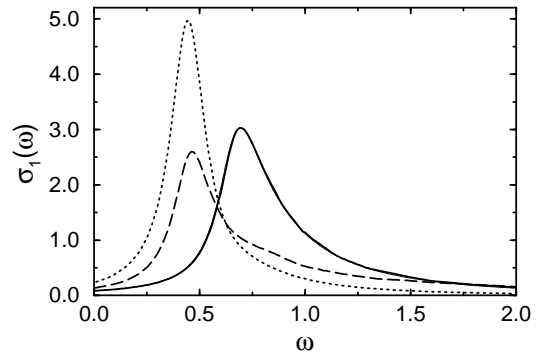}
\caption{\label{fig11}
Comparison of optical spectra calculated with DDMRG 
($N=128$ sites) and field theory ($N = \infty$) for 
$\eta/t=0.1$.  
The dashed line is the DDMRG result for $U=6t$ and $V=3.155t$.
The dotted line is the corresponding field-theoretical result 
($\beta^2=0.327, E_M=0.643t$).
The DDMRG spectrum for $U=4.15t, V=2t$ and the field-theoretical 
spectrum
for $\beta^2=1/2$, $E_M=0.645$ are given by the solid line. 
}
\end{figure}

For $V \leq 2t$ (corresponding to $\beta^2 \geq 1/2$)
one can find parameters so that $\sigma_1^{\text{FT}}(\omega)$ 
perfectly fits
the numerical data over a wide frequency range.
For instance, in Fig.~\ref{fig11} no difference is visible between
the DDMRG spectrum for $U=4.15t$ and $V=2t$ and the fitted
field-theoretical spectrum up to $\omega =2t \approx 3 E_M$.
For $V > 2t$, however, discrepancies between DDMRG and field-theory
results appear and grow progressively stronger as $V$ 
increases.
It is no longer possible to find parameters $\beta$ and $A$ to
reproduce the DDMRG spectra over a significant frequency range above
$\omega=E_{\text{opt}}$. 
Instead $\beta$ and $A$ are set by the optical gap and the total
spectral weight. This yields the values of $\beta$ shown
in parenthesis in Table~\ref{table1}.
As an example, one see in Fig.~\ref{fig11} that the field-theoretical 
spectrum differs significantly from the DDMRG result for
$U=6t$ and $V=3.115t$, although optical gap, Mott gap, and total 
spectral weight are identical for both spectra.
The DDMRG result shows that there is substantial optical weight
both at $E_{\text{opt}}=0.445t$ and above the Mott gap $E_M=0.643t$
while, according to field theory,~\cite{EGJ} 
for a ratio $E_{\text{opt}}/E_M \approx 0.7$ the optical conductivity 
should be dominated by an excitonic peak at $\omega=E_{\text{opt}}$ 
with very little  weight in the holon-antiholon band above the Mott gap 
$E_M$ and in the exciton-exciton continuum above $\omega \approx 1.7 
E_{\text{opt}} \approx 1.1 t$. 

This disagreement between field theory and DDMRG results       
is not due to a failure of either method in the excitonic regime.
It has been shown that field theory and DDMRG calculations
for excitons agree very well in the extended Hubbard model
with next-nearest-neighbor repulsion.~\cite{EGJ}
The problem is that the field-theory approach assumes that the
low-energy optical excitations are made of two 
elementary charge excitations.
In the extended Hubbard model~(\ref{hamiltonian}), however,
the conditions $E_M \ll 4t$ and $V > 2t$ are satisfied only close to the
phase boundary $V_c =U/2$, where low-energy excitations are CDW droplets
made of many elementary charge excitations.
For instance, the ionicity of the $1B_u^-$ state becomes significantly
larger than 1 as $V$ increases above $2t$ as seen in Table~\ref{table1}.
Therefore, the field-theory approach is not applicable 
to the lattice model~(\ref{hamiltonian}) with $V > 2t$ even
in the limit of a small Mott gap.
Moreover, the extended Hubbard model~(\ref{hamiltonian}) cannot 
describe a Mott insulator with a small gap and an exciton in the 
optical spectrum for any parameters $U$ and $V$.

\section{Conclusion \label{sec:conclusion}}

\begin{figure}
\includegraphics[width=8cm]{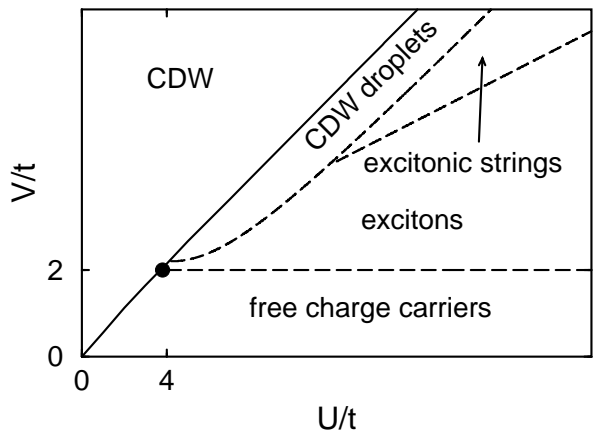}
\caption{\label{fig12}
Schematic representation of the different regions in 
the $(U,V)$ parameter space where a particular optical excitation 
dominates the low-energy optical spectrum of the Mott insulating phase. 
The solid line is the boundary between the CDW and Mott phases.
}
\end{figure}

I have investigated the linear (one-photon) optical excitations of a 
one-dimensional Mott insulator, the half-filled extended Hubbard model, 
using DMRG methods.
Four types of optically excited states have been found: pairs of free 
(unbound) charge excitation, 
excitons, excitonic strings, and CDW droplets.
Correspondingly, there are four different regimes in the model 
parameter space $(U,V)$ depending on the nature of the low-energy 
optical excitations.
They are shown in the schematic ``phase diagram'' of Fig.~\ref{fig12}.
Note that only the $V=2t$ line separating the regime of free excitations
from that of bound excitations represents a sharp transition.
The other dashed lines represent smooth crossover from one regime 
to another.
In each regime one observes optical spectra with distinct features.
In all cases, 
optical excitations are made of an even number of elementary 
excitations carrying opposite 
charges in the lower and upper Hubbard bands.
The different types of excitations and optical spectra found
in this system result from the residual interactions between
these elementary excitations.

For $V \leq 2t$ the low-energy optical excitations are made of two
unbound elementary charge excitations.  
They give rise to a single continuous band in the optical spectrum,
starting at the Mott gap $E_M$, which thus equals the optical gap 
$E_{\text{opt}}$. 
The optical conductivity vanishes smoothly as 
$\sqrt{\omega-E_{\text{opt}}}$ at the threshold $E_{\text{opt}}$,
except for $V=2t$, where  it diverges as 
$1/\sqrt{\omega-E_{\text{opt}}}$.
For $V > 2t \ ( \Rightarrow U \agt 4t)$ but $V < U/3 + O(t)$
and $U-2V \agt t$, the lowest optical excitation
is an exciton (a neutral excitation made of two bound elementary 
charge excitations) 
at an energy $E_{\text{opt}}$ lower than the Mott gap $E_M$.
This exciton gives rise to an isolated $\delta$-peak in the optical 
conductivity at $\omega = E_{\text{opt}}$.
Furthermore, one still finds a continuous band starting
at the Mott gap $E_M$ due to free charge excitations.
For $V > U/3 + O(t)$ and $V > 2t$ but $U-2V \agt t$ 
($ \Rightarrow U \agt 12 t$)
the low-energy optical excitations
are excitonic strings consisting of $n_{\text{exc}} \geq 1$ excitons
with energy $E(n_{\text{exc}}) \approx E_{\text{opt}} + 
(n_{\text{exc}}-1) (U-2V)$.
The spectrum consists of several isolated $\delta$-peaks at
$\omega = E(n_{\text{exc}}) < E_M$.
In this regime one still observes a very weak band of free charge 
excitations starting at $E_M$ in the spectrum.   
Finally, close to the CDW phase boundary ($U-2V \alt t$) for $V > 2t$
the low-energy excitations are CDW droplets and give rise to a
broad band starting below the Mott gap $E_M$.

As long as optical excitations are made of a pair of (bound or unbound)
elementary charge excitations (i.e, the excitation ionicity is 
$I \alt 1$), 
the optical spectra calculated numerically with DDMRG agree perfectly 
with the analytical results obtained with a strong-coupling analysis or
with field-theoretical methods.
This agreement confirms the accuracy and the power of the DDMRG method 
for calculating dynamical spectra in the thermodynamic limit.
It also confirms the wide range of validity of both analytical 
approaches.

Some results presented here suggest further investigations.  
First, in the extended Hubbard model~(\ref{hamiltonian}) 
no exciton exists in the regime of a small Mott gap ($E_M \alt t$),
which is relevant for some real materials such as conjugated 
polymers.~\cite{kiess,sari} 
It is believed that an electron-electron interaction with a longer 
range~\cite{EGJ,gallagher,barford}
or a lattice dimerization~\cite{shuai,joerg2} 
can lead to the formation of excitons in systems with small gaps.
However, the precise nature of the optical excitations in such systems
is still controversial.~\cite{robert}
The approach used here for the extended Hubbard model
will enable us to determine the optical properties of these systems
reliably.
Second, excitonic strings appear in the \textit{linear}
optical conductivity spectrum for strong interaction $U,V \gg t$
because of the weak hybridization of excitonic strings
with different sizes $n_{\text{exc}}$.
Experimentally, excitonic strings have been observed in the 
\textit{non-linear} optical absorption only.~\cite{mazumdar}  
It would be desirable to check if excitonic strings can be found in 
the linear optical absorption of materials which are believed
to be large-gap one-dimensional Mott insulators such as
Cu oxides and Ni halides.~\cite{hasan,kishida,mizuno}      
Last, there is a clear boundary between free and bound 
excitations in the low-energy optical spectrum at $V=2t$.
As discussed in Ref.~\onlinecite{jeckel2} 
the nature of the low-energy charge 
excitation seems to be correlated with the order of the transition from
the Mott insulating phase to the CDW insulating phase.
It is likely that the tricritical point where 
the transition changes from continuous to first order is located
precisely on the line $V=2t$.
This suggests the existence of a hidden symmetry in the charge sector 
of the extended Hubbard model~(\ref{hamiltonian}) at $V=2t$. 
It would be interesting to investigate this feature further.

\begin{acknowledgments}
I gratefully acknowledge helpful discussions with R. Bursill, 
F.~Essler, and S.~Pleutin and I thank F.~Gebhard for his support and 
many stimulating conversations.
\end{acknowledgments}

\appendix*
\section{}

In the original implementation of the charge-conjuga\-tion and 
spin-flip symmetries for DMRG calculations,~\cite{ramasesha} 
an explicit matrix representation of the superblock Ham\-iltonian is 
built.
This matrix can be projected onto a symmetry subspace
with chosen parities $P_{c}$ and $P_{s}$,
which allows one to compute eigenstates of this symmetry 
and reduces the computer memory and CPU time required.
In an efficient implementation of DMRG, however,
an explicit representation
of the superblock Hamiltonian should not be constructed
(at least for quasi-one-dimensional systems with only short-range 
interactions). 
A representation in terms of tensor products of matrices uses
much less memory and is also much faster.~\cite{reinhard}
Projecting this representation onto a symmetry subspace
slows down the program considerably.
Therefore, instead of a projection, I use an exact 
diagonalization technique~\cite{xenophon}
to shift the chosen symmetry subspace to lower energy.

Let $\hat{P}_{c}$ and $\hat{P}_{s}$ be the charge-conjugation and 
spin-flip operators for the full lattice
with eigenvalues $P_{c} = \pm 1$ and $P_{s} = \pm 1$.
As $\hat{P}_{c}$ and $\hat{P}_{s}$ commute with the
Hamiltonian $\hat{H}$, the operator  
\begin{equation}
\hat{H}' = \hat{H} - \lambda_{c} \hat{P}_{c}
- \lambda_{s} \hat{P}_{s}
\end{equation}
has the same eigenstates as $\hat{H}$ but its eigenvalues
are shifted, $E'_n = E_n \pm \lambda_{c} \pm \lambda_{s}$,
where the signs $\pm$ are given by the eigenstate parities 
$P_{c}$ and $P_{s}$.  
It is obvious that the lowest eigenstates of $\hat{H}'$
lies in the symmetry subspace with 
$P_{c} = \lambda_{c}/|\lambda_{c}|$
and $P_{s} = \lambda_{s}/|\lambda_{s}|$ provided
$|\lambda_{c}|$ and $|\lambda_{s}|$ are large enough.
Therefore, one can simply apply the usual ground-state
DMRG approach to the Hamiltonian $\hat{H}'$ with appropriate
values of $\lambda_{c}$ and $\lambda_{s}$ to obtain
the lowest eigenstates in any symmetry sector.
A similar approach has already been used to shift states with
high total spin $S$ to higher energy in a DMRG 
calculation.~\cite{stephane}
Using the method proposed recently for including a non-abelian 
symmetry group in a DMRG calculation would be a further 
improvement.~\cite{ian}


\vfill


\begin{thebibliography}{99}

\bibitem{kiess} \textit{Conjugated Conducting Polymers}, edited
by H.~Kiess (Springer, Berlin, 1992).

\bibitem{sari} \textit{Primary Photoexcitations in Conjugated Polymers:
Molecular Exciton versus Semiconductor Band Model},
edited by N.S.~Sariciftci (World Scientific, Singapore, 1999).

\bibitem{farges} \textit{Organic Conductors}, edited by J.-P.~Farges
(Marcel Dek\-ker, New York, 1994).

\bibitem{bourbonnais} C. Bourbonnais and D. J\'erome, in
\textit{Advances in Synthetic Metals, Twenty Years of Progress
in Science and Technology}, edited by P. Bernier, S. Lefrant, and
G. Bidan (Elsevier, New York, 1999), pp. 206-301.

\bibitem{hasan} M.Z. Hasan \textit{et al.}, \prl \textbf{88}, 177403
(2002).

\bibitem{kishida} H. Kishida \textit{et al.}, Nature (London) 
\textbf{405}, 929 (2000).

\bibitem{mott} N.F.~Mott, \textit{Metal-Insulator Transitions}, 2nd ed. 
(Taylor \& Francis, London, 1990).

\bibitem{florianbook} F.~Gebhard, \textit{The Mott Metal-Insulator 
Transition} (Sprin\-ger, Berlin, 1997). 

\bibitem{guo} D. Guo \textit{et al.}, \prb \textbf{48}, 1433 (1993). 

\bibitem{diago} 
S. Mazumdar and S.N. Dixit, Synth. Met. \textbf{28}, D463 (1989); 
R.M.~Fye \textit{et al.}, \prb \textbf{44}, 6909 (1991).
J. Favand and F. Mila, \prb \textbf{54}, 10425 (1996).

\bibitem{JGE} E.~Jeckelmann, F.~Gebhard, and F.H.L.~Essler, 
\prl \textbf{85}, 3910 (2000).

\bibitem{controzzi} D. Controzzi, F.H.L. Essler, and A.M. Tsvelik, 
\prl \textbf{86}, 680 (2001); 
in \textit{New Theoretical Approaches to Strongly Correlated Systems}, 
edited by A.M.~Tsvelik (Kluwer, Dordrecht, 2001).

\bibitem{EGJ} F.H.L.~Essler, F.~Gebhard, and E.~Jeckelmann, \prb
\textbf{64}, 125119 (2001). 

\bibitem{florian} F.~Gebhard, K.~Bott, M.~Scheidler, P.~Thomas, and 
S.W.~Koch, Philos.~Mag.~B~\textbf{75}, 47 (1997).

\bibitem{gallagher} F.B. Gallagher and S. Mazumdar, \prb \textbf{56}, 
15025 (1997).

\bibitem{stephan} W.~Stephan and K.~Penc, \prb \textbf{54}, R17269 
(1996).

\bibitem{mizuno} Y. Mizuno, K. Tsutsui, T.~Tohyama, and S. Maekawa,
\prb \textbf{62}  R4769 (2000).

\bibitem{barford} W. Barford, \prb \textbf{65}, 205118 (2002).

\bibitem{steve} S.R.~White, \prl \textbf{69}, 2863 (1992); 
\prb \textbf{48}, 10345 (1993).

\bibitem{dmrgbook} \textit{Density-Matrix Renormalization},
edited by I.~Peschel, X.~Wang, M.~Kaulke, and K.~Hallberg 
(Springer, Berlin, 1999). 

\bibitem{jeckel} E.~Jeckelmann, \prb \textbf{66}, 045114 (2002).

\bibitem{hirsch} J.E.~Hirsch, \prl \textbf{53}, 2327 (1984).

\bibitem{jeckel2} E.~Jeckelmann, cond-mat/0204244 (unpublished).

\bibitem{hubbard} J.~Hubbard, Proc.~R.~Soc.~London,
Ser. A~\textbf{276}, 238 (1963).

\bibitem{lieb} E.H.~Lieb and F.Y.~Wu, \prl \textbf{20}, 1445 (1968).

\bibitem{shuai} Z.~Shuai, S.K.~Pati, W.P.~Su, J.L.~Br\'{e}das, 
and S.~Ramasesha, \prb \textbf{55}, 15368 (1997);
Z.~Shuai, J.L.~Br\'edas, S.K.~Pati,  
and S.~Ramasesha, \prb~\textbf{58}, 15329 (1998).

\bibitem{kancharla} S.S.~Kancharla and C.J.~Bolech,
\prb \textbf{64}, 085119 (2001). 

\bibitem{tomita} N. Tomita and K. Nasu, \prb \textbf{63}, 085107 (2001).

\bibitem{ramasesha} 
S.~Ramasesha, S.K.~Pati, H.R.~Krishnamurthy,
Z.~Shuai, and J.L.~Br\'edas, Phys.~Rev.~B.~\textbf{54}, 7598 (1996).

\bibitem{kuzmany} For an introduction, see
H.Kuzmany, \textit{Solid-State Spectroscopy}
(Springer, Berlin, 1998).

\bibitem{mazumdar} S. Mazumdar \textit{et al.}, Chem. Phys. 
\textbf{104}, 9283 (1996).

\bibitem{pleutin} S. Pleutin, \prb \textbf{61}, 4554 (2000).

\bibitem{joerg} J.~Rissler, F.~Gebhard, H.~B\"{a}ssler, and
P.~Schwerdtfeger, \prb \textbf{64}, 045122 (2001).  

\bibitem{joerg2} J.~Rissler, Ph.D. thesis, University of Marburg, 2001.

\bibitem{robert} M.~Boman and R.J.~Bursill, \prb \textbf{57}, 15167 
(1998); R.J.~Bursill \textit{ibid.} \textbf{63}, 157101 (2001).

\bibitem{reinhard} See R.M.~Noack and S.R.~White, in 
Ref.~\onlinecite{dmrgbook}.

\bibitem{xenophon} X. Zotos, in \textit{Applications of Statistical
and Field Theory Methods to Condensed Matter}, edited by
D. Baeriswyl, A.R. Bishop, and J. Carmelo 
(Plenum Press, New York, 1990). 

\bibitem{stephane} S. Daul and R.M. Noack, \prb \textbf{58}, 2635 
(1998).

\bibitem{ian} I.P.~McCulloch and M. Gul\'{a}csi, 
Europhys. Lett. \textbf{57}, 852 (2002).


\end{thebibliography}
\end{document}